\documentclass{article}

\usepackage[final,sglblindworkshop]{neurips_2026}
\workshoptitle{2nd SIMBIOCHEM Workshop at NeurIPS 2026}

\usepackage[utf8]{inputenc}
\usepackage[T1]{fontenc}
\usepackage{hyperref}
\usepackage{url}
\usepackage{booktabs}
\usepackage{amsfonts}
\usepackage{nicefrac}
\usepackage{microtype}
\usepackage{xcolor}
\usepackage{amsmath}
\usepackage{amssymb}
\usepackage{amsthm}
\usepackage{cleveref}
\usepackage{graphicx}

\theoremstyle{remark}

\crefname{remark}{Remark}{Remarks}\Crefname{remark}{Remark}{Remarks}
\crefname{equation}{Eq.}{Eqs.}\Crefname{equation}{Eq.}{Eqs.}
\crefname{section}{Section}{Sections}\Crefname{section}{Section}{Sections}
\crefname{table}{Table}{Tables}\Crefname{table}{Table}{Tables}
\crefname{figure}{Figure}{Figures}\Crefname{figure}{Figure}{Figures}

\definecolor{editgreen}{RGB}{0,120,0}

\title{The Reference Set Is Not the Answer Set:\\
GEOM-DRUGS Is Missing Low-Energy Conformers on Its Own Energy Surface
}

\author{%
  Alvand Vahediahmar$^{1,4}$ \quad
  Aaron Zweig$^{1}$ \quad
  Mingxuan Zhang$^{1,2}$ \quad
  Elham Azizi$^{1,3,4}$ \\[6pt]
  $^1$Irving Institute for Cancer Dynamics, Columbia University \\
  $^2$Department of Systems Biology, Columbia University \quad \\
  $^3$Department of Computer Science, Columbia University \\
  $^4$Department of Biomedical Engineering, Columbia University \quad
}

\begin{document}

\maketitle

\begin{abstract}
Machine-learned conformer generators are scored on GEOM-DRUGS by proximity to a reference ensemble
that standard benchmark metrics implicitly treat as ground truth. We test whether that finite set is complete
on the GFN2-xTB surface used to construct it. Across five public generators, settled, low-energy
structures remain outside the $0.75$\,\AA{} reference radius after removing reflections, incorrect
stereochemistry, and post-relaxation connectivity changes ($2.3$--$7.0\%$ of submitted conformers,
depending on architecture). In a targeted DFT audit, $107$ of $113$ chemically screened
novel/reference pairs remain outside that radius after full $\omega$B97X-D3BJ re-optimization. A
threshold-free complementary test needs no radius and no energy window at all: on $15.0\%$ $[8.0,
22.0]$ of molecules some generator has already found a chemically intact structure lying below the
\emph{entire} released ensemble. A complementary CREST search reproduces the minimum released in
 $65$ of $69$ molecules, and in $11$ of them it reaches more than
$0.5$\,kcal/mol below the complete released ensemble. 
Augmenting the references with conformers proposed by held-out architectures raises absolute
coverage-precision for every generator tested, by $1.48$--$2.99$ points. Together, these results show that GEOM-DRUGS precision is sensitive not only to generator quality but also to incomplete reference support.
\end{abstract}

\section{Introduction}
\label{sec:intro}

Predicting the 3D shapes a small molecule adopts, its \emph{conformers}, is a standard task in
computational chemistry and a standard benchmark for generative models. The field commonly evaluates on
GEOM-DRUGS \citep{axelrod2022geom}, which supplies for each molecule a reference ensemble obtained by
extensive semi-empirical sampling, and scores a generator by how well its output matches that ensemble under a $0.75$\,\AA{} RMSD threshold. Every published metric on this benchmark treats the
reference ensemble as ground truth, including metrics for  classification tasks. The reference ensembles were produced by a finite search, and if they omit a low-energy basin, a generated conformer in that basin is counted as a precision error. Our goal is to measure this support limitation and make scores easier to interpret. In this paper, we ask whether the released set is sufficiently complete to serve as the positive support of the benchmark.

Benchmark limitations have been studied extensively, where~\citet{northcutt2021labelerrors} find pervasive label
errors in ten standard test sets and show they destabilize model rankings.  Precision/recall decompositions
\citep{sajjadi2018assessing,kynkaanniemi2019improved,naeem2020reliable} note that reference support
determines what can be measured but assume the reference is the target.
\citet{geomdrugs2025revisited} ask whether reference structures are \emph{individually accurate}
where we ask whether the set is \emph{collectively complete}.  Our contributions include:

\begin{enumerate}
\item \textbf{Controlled evidence of missing support} (\Cref{sec:gt}): we run a common-screen audit that
  identifies plausible omitted basins and determines if they can be explained by reflections, stereochemical errors, or altered species.
\item \textbf{An independent search test} (\Cref{sec:crest}): we rerun CREST on released
  structures and recover omitted conformers, including an acylhydrazone rotamer that remains favorable
  after DFT optimization.
\item \textbf{A reproducible reference-augmentation pipeline}: we build a pipeline to screen, validate,
  deduplicate, and augment the GEOM-DRUGS ensembles, are running it across the full set, and will
  release the completed reference data and code.
\end{enumerate}


\section{Setup}
\label{sec:bg}

\paragraph{Metrics.} For generated set $G$ and reference set $R$ with $\delta$ the
automorphism-corrected heavy-atom RMSD after alignment, coverage-recall is
$|\{r\in R:\min_{g\in G}\delta(g,r)<\tau\}|/|R|$ at $\tau=0.75$\,\AA{}.  Average minimum RMSD is
$\mathrm{AMR\text{-}R}=\frac{1}{|R|}\sum_{r\in R}\min_{g\in G}\delta(g,r)$. For precision, the definitions swap the role of
$G$ and $R$. The benchmark fixes the submission size at $2n_{\mathrm{ref}}$. Note that $\delta$ \citep{rdkit} permits proper rotations only, so reflections are excluded. Thus, coverage-precision implicitly assigns every generated conformer farther than $\tau$ from $R$ to the ``unsupported'' side of the metric, irrespective of whether the conformer is physically plausible.

\paragraph{Screening a generated conformer.}
For a generated conformer to count as a valid, successful sample, we apply two sets of
screens, so that we are not counting a model artifact or an unstable structure. The first is the
\emph{basic screen}. We relax the generated conformer with GFN2-xTB \citep{bannwarth2019gfn2} and
measure the RMSD between the relaxed and the unrelaxed geometry. We call this displacement the
conformer's \emph{drift}. Drift measures how far a structure has to travel before it reaches a
minimum, so a small drift indicates that the generated conformer was already inside a low-energy
region rather than on a slope leading into one. We require drift below $0.5$\,\AA{} and call a
conformer that passes \emph{settled}. The second requirement is that the relaxed energy fall within
a window $W$ (the \emph{retention window}) of the lowest energy seen for that molecule, taken from
either the generated pool or the reference set, whichever is lower. We then measure the RMSD from
the generated conformer to every reference conformer; if it exceeds $0.75$\,\AA{} for all of them,
we call the conformer \emph{unreferenced}. Throughout, we use \emph{basin} for a low-energy
conformational region, meaning the set of geometries that relax to the same minimum. The basic
screen therefore isolates conformers that are settled, low in energy, and geometrically novel.

\paragraph{Chemical-identity screens.} The second set of filters establishes not only that these conformers have valid
geometry and low energy, but also that they are not duplicates of existing reference
conformers and do not have incorrect stereochemistry. An unreferenced conformer can be
chemically irrelevant for three reasons that RMSD and energy alone cannot resolve.
(I) \emph{Reflection.} We mirror every reference conformer and measure its RMSD to the
unreferenced conformers; if any mirrored structure falls within $0.75$\,\AA{} of an
unreferenced conformer, we discard that conformer. (II) \emph{Stereochemistry.} We check
the three-dimensional arrangement of the molecule to determine how its substituents are oriented in space, which distinguishes, for example, left-handed from right-handed forms.
To remain unreferenced, a conformer must match the reference assignment at every
stereocenter and every stereogenic bond. (III) \emph{Post-relaxation connectivity.}
After relaxation, the molecular graph must be unchanged: the conformer must retain the
same heavy-atom bonds and the same hydrogen counts as before relaxation, meaning that no
bonds were broken or formed during the process. Applying the first two filters, we find
that graph-conditioned models rarely fail while Cartesian models fail often; together,
the filters remove roughly half of the candidates. \Cref{tab:main} reports the result after the full screen. Between $2.3$ and $7.0\%$ of all
submitted conformers survive it. Conditioning on settling, that is, restricting the denominator to
the conformers that pass the drift test rather than counting all submitted conformers, the figure is
$8.6$--$16.8\%$. Before the chemical screens the same two quantities read $4.6$--$19.7\%$ and
$18.6$--$44.2\%$. A large share of what a naive reading of this criterion calls a missing basin is
therefore a generator's stereochemistry error rather than a gap in the benchmark.

None of the three constants is ours to choose. The $0.75$\,\AA{} clustering radius is the
scoring threshold defined by the benchmark itself. The $0.5$\,\AA{} drift tolerance is smaller than
that radius, so a settled conformer is required to move less than the distance at which two
structures would already be counted as the same basin, which makes it the more stringent of the two.
The $6$\,kcal/mol window is the retention criterion used to construct the reference set. We report
the sensitivity of our results to all three in \Cref{fig:sens}.

\section{The reference ensemble is incomplete}
\label{sec:gt}

\Cref{tab:main} is measured on one common cohort so that the five architectures are directly
comparable. We sampled $500$ test molecules (half of GEOM-DRUGS test-set) and kept the $384$ on which all five checkpoints
produced a usable pool, so the cohort is the intersection of the five pools rather than a filter
applied to the molecules themselves. Each pool is capped at $25$ conformers per molecule, and every
row uses the same constants and the same interval procedure.

\begin{table}[t]
\centering
\small
\caption{For $n=384$ molecules, pools capped at $25$
conformers, $W=6$\,kcal/mol, drift $\le 0.5$\,\AA{}, $0.75$\,\AA{} radius. Entries are conformer
fractions with $95\%$ molecule-clustered bootstrap intervals. Columns 3 and 4 are conditional on
settling, column 5 is over all submitted conformers. \emph{Screened} requires no reflection match,
correct defined stereochemistry, and an unchanged post-relaxation bond and H-count pattern.}
\label{tab:main}
\resizebox{\textwidth}{!}{%
\begin{tabular}{lllll}
\toprule
Generator & xTB-settled & Unref.\ $\mid$ settled & \textbf{Unref.\ $\mid$ settled, screened} &
  \textbf{Unreferenced endpoints, screened} \\
\midrule
GeoMol & $0.3195$ $[0.2856, 0.3534]$ & $0.2032$ $[0.1673, 0.2406]$ & $0.1219$ $[0.0922, 0.1535]$ & $0.0226$ $[0.0164, 0.0297]$ \\
Torsional diffusion & $0.4910$ $[0.4568, 0.5257]$ & $0.2057$ $[0.1785, 0.2344]$ & $0.1175$ $[0.0952, 0.1414]$ & $0.0395$ $[0.0322, 0.0473]$ \\
MCF & $0.6756$ $[0.6479, 0.7031]$ & $0.3388$ $[0.3100, 0.3687]$ & $0.0962$ $[0.0800, 0.1137]$ & $0.0545$ $[0.0458, 0.0641]$ \\
ET-Flow & $\mathbf{0.8674}$ $[0.8471, 0.8866]$ & $0.1858$ $[0.1643, 0.2085]$ & $0.0855$ $[0.0707, 0.1014]$ & $\mathbf{0.0701}$ $[0.0582, 0.0829]$ \\
GeoDiff & $0.2109$ $[0.1879, 0.2348]$ & $\mathbf{0.4415}$ $[0.3997, 0.4839]$ & $\mathbf{0.1683}$ $[0.1366, 0.2014]$ & $0.0292$ $[0.0227, 0.0365]$ \\
\bottomrule
\end{tabular}}
\end{table}

We ran nine controls against this result, each described in full in \Cref{tab:controls}.
The strongest is a frozen DFT audit on a separate cohort: after screening for stereochemistry,
topology and charge, $113$ pairs from $108$ molecules remain, and $107$ of them are still separated
by more than the $0.75$\,\AA{} scoring radius after $\omega$B97X-D3BJ/def2-SVP optimization. This
cohort was selected to favor large separations and agreement across architectures, so it establishes
that robust examples exist rather than estimating an average survival rate, and without DFT Hessians
these are optimization endpoints rather than frequency-certified minima. Each comparison is also
pairwise, between a candidate and the single reference conformer nearest to it; it does not show
that the candidate lies below the minimum of the entire released ensemble. The test below addresses
that question directly.

Screened unreferenced conformers sit close to the boundary that excludes them, at a median RMSD of
$1.005$\,\AA{} from the nearest reference conformer against a scoring radius of $0.75$\,\AA{}. Under
an explicit atom-level torsion mapping to that nearest conformer, $47.2\%$ of them differ from it in
only one or two torsions, with ring conformations essentially unchanged (\Cref{app:controls}). The
omissions are therefore local rearrangements of structures the reference set already contains rather
than unrelated geometries. They are also not spread evenly across the dataset: weighting each
conformer by a room-temperature Boltzmann factor of its electronic energy places most of the missing
population in a small subset of molecules, though this weighting ignores entropy, degeneracy,
solvation and basin volume and we report it only as a diagnostic (\Cref{app:boltzmann}). One
plausible cause is the cost of the search itself. \citet{axelrod2022geom} report that some molecules
were dropped from GEOM entirely because CREST did not converge within several days on $40$ cores.
What we observe is consistent with a milder form of the same constraint, in which the search
completed but did not exhaust the basins accessible within a molecule.

\paragraph{A minimum-energy test independent of the RMSD radius.} For each molecule we
compare the lowest relaxed energy found anywhere in a generator's pool against the lowest energy in
the entire reference ensemble. This comparison uses no RMSD radius and no retention window. It
requires only a numerical energy margin and a check that the winning structure is the intended
molecule. At least one generator produces a pool minimum below the reference minimum on $22.0\%$ of
molecules (\Cref{tab:globalmin}). Requiring the winning structure to retain the molecule's intended
local valence and hydrogen-count signature, which prevents a chemically altered structure from
winning, reduces this to $\mathbf{15.0\%}$ $[8.0, 22.0]$. A global minimum is not removed by any
pruning step in the reference construction workflow, and CREST restarts its sampling whenever it
finds a lower-energy structure, with \citet{axelrod2022geom} reporting no limit on the number of
restarts. Each of these cases therefore involves a structure the reference search should in
principle have reached.

Taken together, this section screens generated conformers for every artifact we can test for
and finds that a residue of settled, low-energy, chemically intact structures still lies outside the
reference ensemble.

\section{An independent CREST search}
\label{sec:crest}

The audit in \Cref{sec:gt} showed numbers of molecules with conformers absent from the reference ensemble. We therefore run the search program that built GEOM-DRUGS to see whether it recovers those structures on its own.

\paragraph{Search protocol.}
CREST \citep{pracht2020crest,pracht2024crest} is the search program that was used to build the GEOM-DRUGS reference ensembles. We ran CREST on 69 test molecules that we chose at random. We used CREST version 2.9, which is the version used to build GEOM. Each search started from one released reference conformer, at the same formal charge as the released entry. For every molecule we ran two searches. The first one used CREST's default settings. The second one used four times the default metadynamics length (\texttt{-mdlen x4}). The metadynamics length sets how long CREST keeps searching before it stops, so making it four times longer gives the search a larger budget.
We used the same CREST binary and the same formal charge in both searches, so the metadynamics
length is the only setting that changes between them. \Cref{tab:crestfull} and \Cref{tab:crestkeys}
give the full results for each molecule.

CREST keeps a conformer only if its relaxed energy is inside its own energy window $E_{\mathrm{win}}$, which is
measured from the lowest energy conformer that the search found, and the default value of
$E_{\mathrm{win}}$ is $6$ kcal/mol. A larger or smaller ensemble therefore tells us about this
setting and not about how much of the conformational space the search covered. We report two other
quantities for each molecule. The first one is where the lowest energy conformer of each search
sits compared to the released reference minimum. The second one is how much of each CREST ensemble
is further than $0.75$\,\AA{} from every released reference conformer, using the scoring radius
defined in \Cref{sec:bg}.

\paragraph{Reproducibility check.}
We first checked whether the default search finds the same minimum as the released reference
ensemble, because this tells us whether the two searches can be compared on the same energy scale.
On 65 of the 69 molecules the default search matches the released reference minimum within
$0.5$ kcal/mol. We fixed this $0.5$ kcal/mol margin before we ran the comparison. If we use a much
stricter margin of $10^{-3}$ kcal/mol, then 60 of the 69 molecules still match, and 57 of those 60
match exactly. This agreement supports treating the energies from the two searches as comparable
numbers. We also checked that the CREST conformers did not turn into a different molecule during
relaxation. We relaxed every conformer from both searches, which is more than 35{,}000 conformers
in total. All of them passed the topology screen described in \Cref{sec:bg}, which tests that the
heavy atom bonds and the hydrogen counts stay the same after relaxation.

\paragraph{Results.}
On 11 of the 69 molecules the larger budget search finds a conformer that is more than
$0.5$ kcal/mol below the lowest energy conformer of the released ensemble. This is the same
$0.5$ kcal/mol margin that we used above. On 7 of these 11 molecules the default search had already
matched the released minimum exactly. This pattern fits the idea that the larger search budget is
what finds the lower energy conformers. We ran only one search per setting for each molecule, so a
single run cannot rule out other explanations. \Cref{tab:crestfull} lists the outcome for each
molecule and includes several other examples besides the one that we describe next.

\begin{figure}[t]
\centering
\includegraphics[width=0.70\textwidth]{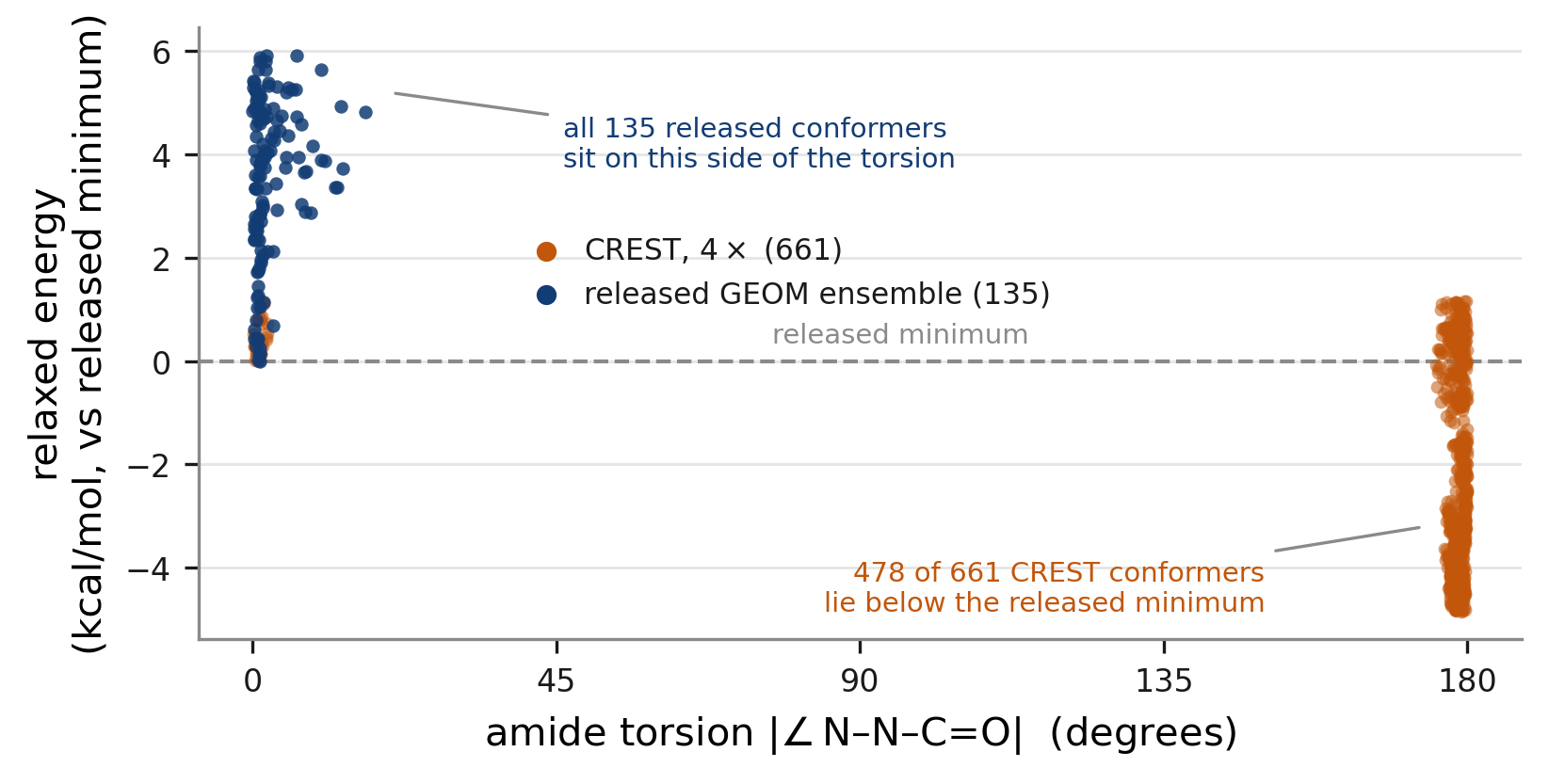}
\caption{Molecule L1. Relaxed GFN2-xTB energy against the amide torsion, zeroed at the released
ensemble's minimum. The released ensemble (dark blue) occupies one rotamer entirely; CREST (dark
orange), started from a released conformer and topology- and stereochemistry-identical to it,
populates both.}
\label{fig:l1}
\end{figure}

\paragraph{Molecule L1.}
\Cref{fig:l1} shows the clearest example, which is molecule L1, an acylhydrazone. Its SMILES
string is given in \Cref{tab:crestkeys}. One possible explanation for the gap on this molecule is
that CREST searched a different chemical species than the released ensemble, for example a
different protomer or a different configurational isomer. If this were true, then the gap would not
mean that the search of the same conformational space was incomplete. We tested this directly. The
CREST search for L1 started from released reference conformer 1 at the released formal charge. We
assigned the stereochemistry of every conformer from its own 3D coordinates, and we obtained the
same isomeric SMILES string for all 135 released conformers and all 661 CREST conformers. The
compound, its configuration and its protonation state are therefore the same in both ensembles.

What differs between the two ensembles is the rotamer that they occupy, which means the side
of one rotatable bond that the conformer sits on. Here that bond is the amide torsion. All 135
released conformers sit on one side of this torsion. Out of the 661 CREST conformers, 612 sit on
the other side. There are 478 CREST conformers with an energy below the released reference minimum,
and all 478 of them sit on this second side, which the released ensemble does not cover. The other
49 CREST conformers sit on the same side as the released ensemble, and their lowest energy matches
the released reference minimum within $0.001$ kcal/mol. This gives an internal control, because the
two searches agree almost exactly on the side that both of them cover. The whole $4.85$ kcal/mol
gap between the two minima therefore comes from this one rotation, which the released search did
not cross.

\begin{figure}[htbp]
\centering
\includegraphics[width=\textwidth]{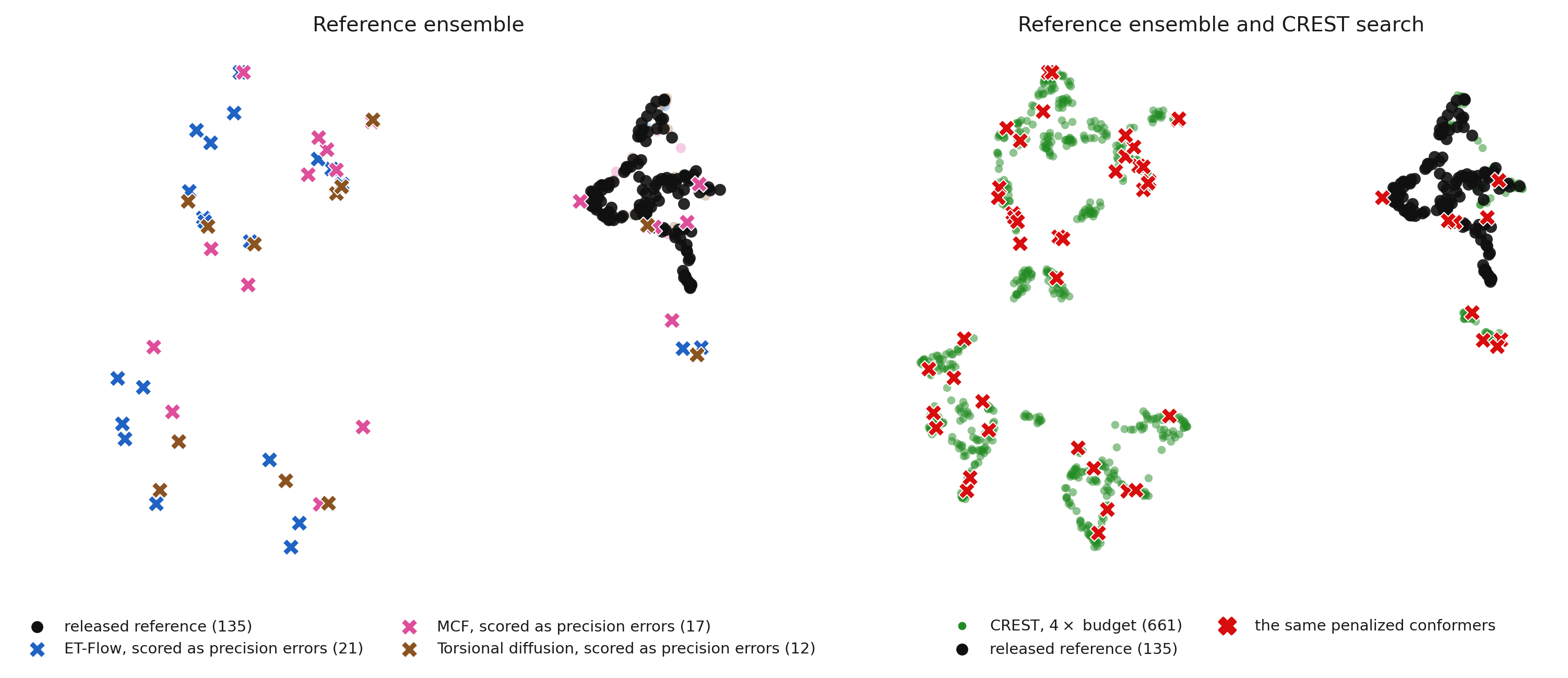}
\caption{Molecule L1. \emph{Left}: the released reference
(black) and the conformers three architectures were scored wrong for (ET-Flow blue, MCF pink,
torsional diffusion brown) form clusters isolated from the reference. \emph{Right}: CREST's $661$
conformers (dark green) land inside those same orphan clusters, overlapping the structures the
benchmark counted as errors (red).}
\label{fig:crest-umap}
\end{figure}

All of these energy comparisons use GFN2-xTB, which is a semi-empirical method. Its absolute
accuracy on a rotational barrier of this kind is not high enough for the $4.85$ kcal/mol figure to
be read as a converged number, and we treat that as a limitation of the method rather than of the
observation, because the comparison is made entirely within one method and does not depend on how
accurate that method is in absolute terms. We then checked whether the preference for the
CREST-side rotamer holds at a higher level of theory. We re-optimized 23 conformers with
$\omega$B97X-D3BJ/def2-SVP density functional theory. None of the 23 moved to the other rotamer
during re-optimization. At this level the rotamer of the released ensemble is $1.91$ kcal/mol higher
in energy than the rotamer that CREST reaches, against $4.85$ kcal/mol on GFN2-xTB, so the cheaper
method overstates the gap by $2.9$ kcal/mol while preserving its sign. \Cref{fig:crest-umap} shows
where the two ensembles sit relative to the conformers that three generator architectures were
penalized for.

\section{Exploratory comparison to protein-bound ligand conformers}
\label{sec:pdb}

\Cref{sec:gt} and \Cref{sec:crest} both judge completeness on the gas-phase surface the
benchmark was built on. Here we ask a different question: does the released ensemble contain the
conformer a ligand actually adopts inside a protein pocket? A bound conformer can be strained, so a
gas-phase ensemble is under no obligation to contain it, and we present this section as exploratory
for that reason.

\begin{figure}[t]
\centering
\includegraphics[width=0.88\textwidth]{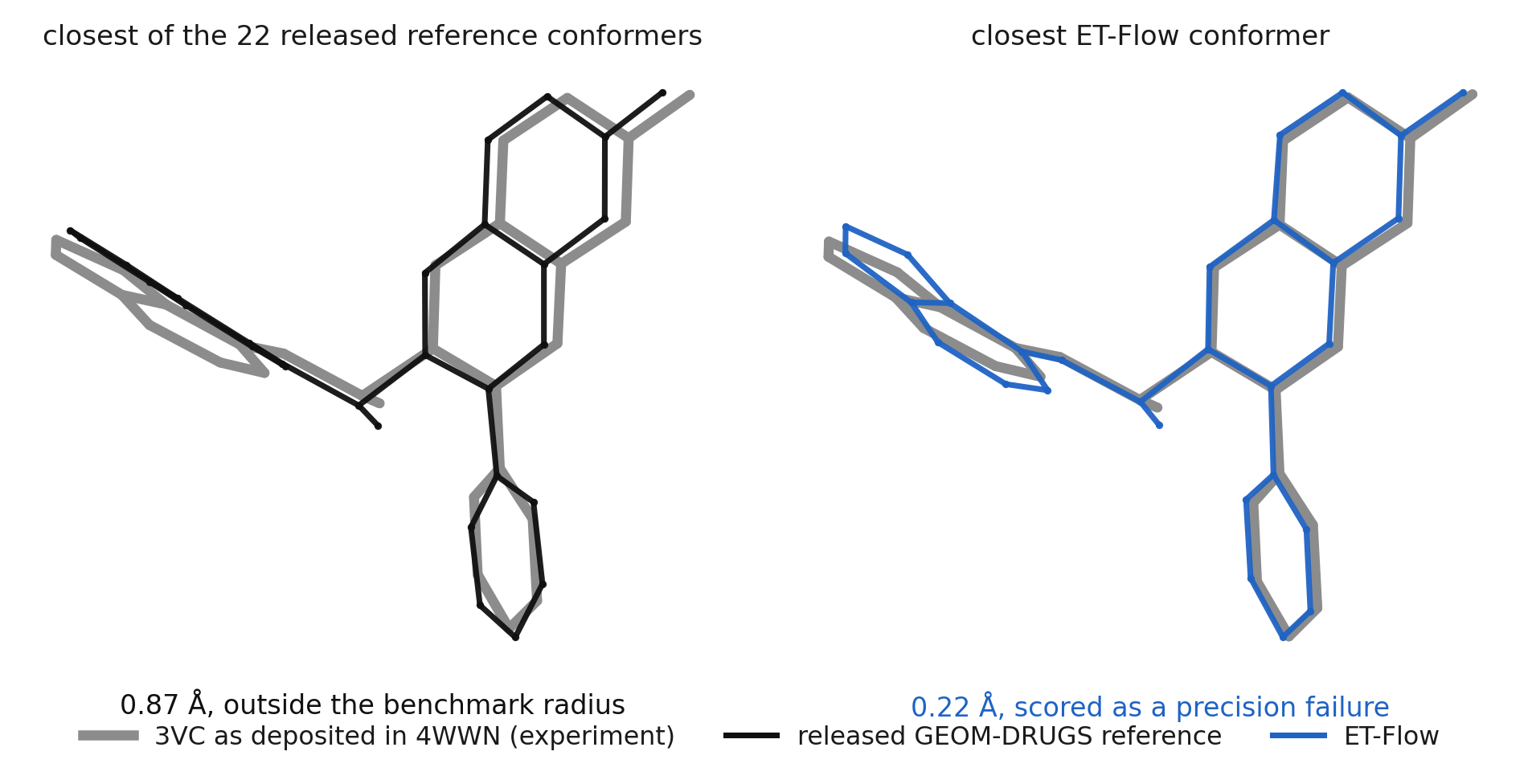}
\caption{Ligand 3VC (PDB 4WWN), projected on the crystal geometry's own principal plane. \emph{Left}:
the closest of $22$ released reference conformers (black) sits $0.872$\,\AA{} from the deposited
geometry (grey), outside the benchmark's $0.75$\,\AA{} radius. \emph{Right}: the closest ET-Flow
conformer (blue) sits $0.223$\,\AA{} away, yet is scored as a precision failure. Both are
automorphism-corrected RMSDs to the same experimental target; direct scoring additionally places
that ET-Flow conformer $0.847$\,\AA{} from its nearest released reference.}
\label{fig:3vc}
\end{figure}

We screened the PDB Chemical Component Dictionary for ligands that also appear in GEOM-DRUGS
and measured three automorphism-corrected heavy-atom RMSDs for each match: from the experimental
conformer to the nearest released reference conformer, from the experimental conformer to the
nearest generated conformer, and from that generated conformer to every released reference
conformer. Only the third determines how the benchmark scores it. Two matches pass all three
measurements together with an exact stereochemistry check against the GEOM-DRUGS entry
(\Cref{tab:pdb}); \Cref{fig:3vc} shows the clearer of them. In both, a generated conformer
reproduces the deposited geometry more closely than any released reference conformer does and is
nevertheless counted as a precision error. These are two existence proofs, not a rate: the screen is
small, the generated pools are far deeper than the references, and the pocket changes the energy
landscape. \Cref{app:pdb} gives the screen and the checks in full.

\begin{table}[t]
\centering
\small
\caption{Two exact-stereochemistry PDB examples, as automorphism-corrected heavy-atom RMSDs.
$d(R,P)$ and $d(g,P)$ measure the released ensemble and the selected generated conformer against
the deposited geometry. Only $d(g,R)$ enters the benchmark score, so a conformer $0.223$\,\AA{} from
the experimental geometry is still counted as a precision error once it lies beyond $0.75$\,\AA{}
of every reference conformer. Ligand codes are PDB Chemical Component Dictionary IDs.}

\label{tab:pdb}
\begin{tabular}{llrrrrl}
\toprule
Ligand & PDB & $n_{\mathrm{ref}}$ & $d(R,P)$ (\AA) & $d(g,P)$ (\AA) & $d(g,R)$ (\AA) & generator \\
\midrule
\textbf{3VC} & 4WWN & $22$ & $\mathbf{0.872}$ & $\mathbf{0.223}$ & $\mathbf{0.847}$ & ET-Flow \\
\textbf{8DE} & 7W7F & $55$ & $\mathbf{0.849}$ & $\mathbf{0.463}$ & $\mathbf{0.772}$ & ET-Flow \\
\bottomrule
\end{tabular}
\end{table}

\section{Sensitivity to augmenting the reference set}
\label{sec:corrected}

\paragraph{Augmented reference set construction.}
We ran a preliminary sensitivity analysis that asks how much the scores change when we add
model-generated conformers to the released reference ensembles. We selected the conformers to add
with the same low-energy conformer logic as \Cref{sec:gt}, but with a wider retention window of
$W=10$ kcal/mol instead of the $W=6$ kcal/mol used elsewhere in this paper, which admits more
candidates. We start from the pool conformers that settle under relaxation and that are at least
$0.75$\,\AA{} away from every released reference conformer. We cluster these conformers at
$0.75$\,\AA{} and keep one conformer per cluster, so that near duplicates are not added twice. We do
this for four generator architectures at matched pool depth and we pool the result. The output is a
candidate key, which is the released reference ensemble together with one conformer for each basin
that the released ensemble does not cover. 

This key is an audit instrument rather than a replacement ground truth: it uses the wider window,
its energy anchors depend on which source a conformer came from, and its added conformers are
generator proposals that have not passed the post-relaxation topology screen of \Cref{sec:gt}.

\paragraph{Leave-one-out scoring.}
We score each architecture against a leave-one-out key that is corrected using only the
conformers proposed by the other four architectures, so no model is ever scored against a key that
contains its own conformers. Under this design the coverage-precision of every architecture rises,
and the $95\%$ interval for every architecture excludes zero. \Cref{tab:looarch} reports these
results on the same 96 molecules for every row. A key built this way inherits the same defect that
\Cref{sec:gt} measures, because some proposed conformers are reflections of reference conformers or
have the wrong stereochemistry. We therefore rebuild the key after the reflection and stereochemistry
screens and report both versions. 
The screened-key column is the sensitivity result we interpret, and the unscreened column shows what
the two screens remove. Each row is scored against a key built from a different set of four
architectures, so the rows measure sensitivity within an architecture and cannot be read as a
ranking across them.

\paragraph{The change in recall.}
The recall column is not a loss of performance. Adding candidate conformers to the key
enlarges the set a method is asked to find, so recall falls even when nothing about the method has
changed. Our claim is therefore limited to what this design supports: absolute coverage-precision
depends on how complete the reference set is, not on any reordering of the published methods.
Settling the question requires a single key built by an independent search and scored by every
architecture on the same molecules.

\begin{table}[t]
\centering
\small
\caption{Leave-one-out reference correction, each architecture held out, $n=96$ molecules paired,
molecule-clustered bootstraps; each row's key is corrected using only the \emph{other four}.
\emph{screened} rebuilds that key after reflection and stereochemistry checks.}
\label{tab:looarch}
\resizebox{\textwidth}{!}{%
\begin{tabular}{lrrrl}
\toprule
Held-out architecture & COV-P before & $\Delta$ unscreened key & $\mathbf{\Delta}$ \textbf{\textit{screened key}} & $\Delta$ recall (unscreened) \\
\midrule
ET-Flow & $78.13$ & $+6.46$ $[+4.82, +8.23]$ & $\mathbf{+2.91}$ $[+1.73, +4.27]$ & $-7.84$ $[-10.05, -5.81]$ \\
Torsional diffusion & $57.21$ & $+5.88$ $[+4.40, +7.48]$ & $\mathbf{+2.99}$ $[+1.90, +4.26]$ & $-6.72$ $[-8.76, -4.88]$ \\
MCF & $53.39$ & $+8.59$ $[+6.56, +10.78]$ & $\mathbf{+2.75}$ $[+1.72, +4.01]$ & $-4.22$ $[-5.93, -2.67]$ \\
GeoMol & $44.30$ & $+3.41$ $[+2.18, +4.89]$ & $\mathbf{+1.48}$ $[+0.68, +2.46]$ & $-4.85$ $[-6.60, -3.26]$ \\
GeoDiff & $21.82$ & $+6.25$ $[+4.29, +8.37]$ & $\mathbf{+1.85}$ $[+0.97, +2.88]$ & $-0.51$ $[-1.81, +0.74]$ \\
\bottomrule
\end{tabular}}
\end{table}

\section{Limitations}
\label{sec:limits}

\paragraph{DFT audit.}
The full DFT audit is a targeted study and not a random sample. It contains 113 pairs that passed
our chemical screens, drawn from 108 molecules, where each pair is one candidate conformer together
with its nearest released reference conformer. Of these 113 pairs, 107 stay outside the
$0.75$\,\AA{} scoring radius after full re-optimization. We chose the pairs rather than sampling
them at random, and we gave priority to clearly separated candidates that several architectures had
proposed. This design shows that such cases exist and survive DFT, but it does not estimate what
fraction of all candidate conformers would survive DFT. We also computed no DFT Hessians, so we
describe the results as optimization endpoints and not as confirmed minima. The audit does not
validate GFN2-xTB energies either. The single point comparison in \Cref{app:trust} gives a Spearman
rank correlation of $0.8762$, which says only that GFN2-xTB orders the conformers within one
molecule the way DFT does, and says nothing about the accuracy of those energies in general.

\paragraph{Sample sizes.}
The sample sizes in this paper are set by compute cost rather than by design, and one CREST molecule
costs about 26 core-hours across both settings. The minimum-energy test uses 100 molecules, of which
81 enter the paired comparison, against a test split of about $1{,}000$ molecules. The Hessian check
uses 216 conformers. The DFT work uses 17 to 36 conformers for the single point comparison and 113
pairs for the full re-optimization. The CREST campaign uses 69 molecules, and 28 of those carry the
two-sided data that the basin completeness rate needs. We tested one CREST budget multiplier, four
times the default metadynamics length, so we did not sweep the budget. This is a pilot study at the
scale of the benchmark it comments on, so we report an interval with every number. Every interval is
a bootstrap interval computed by resampling molecules, and \Cref{app:detail} describes how we
compute it.

\paragraph{Energy criterion.}
Our criterion is itself a model of the physics and not a direct measurement. A neural network
potential used as a cross-check agrees with GFN2-xTB at a Cohen's kappa of $\kappa=0.656$, where
kappa measures agreement after chance agreement is removed, and this agreement degrades for strained
conformers. We sweep the three constants of the criterion, the RMSD radius, the drift tolerance and
the retention window, but not the choice of semi-empirical method. The criterion also describes a
single molecule in the gas phase, so it should not be applied to conformers whose shape an
environment holds in place, such as a ligand inside a protein pocket. \Cref{app:trust} reports what
it returns for those conformers. The topology screen is a geometric test. It compares the bonds
perceived from the 3D coordinates, including bonds to explicit hydrogens, against the intended
template graph, so it detects that the bond pattern changed but not why.

\paragraph{Scope of the cohorts.}
The 384-molecule cohort is the intersection of five cached generator pools drawn from 500 randomly
sampled test molecules. Its intervals describe that cohort and do not automatically generalize to
the full GEOM-DRUGS test split. The 69 CREST molecules were also drawn at random, and we have not
yet identified which properties of a molecule predict when a CREST search reaches below the released
reference ensemble. The candidate keys of \Cref{sec:corrected} inherit the priors of the generators
that proposed them, and each held-out architecture is scored against a different key. The definitive
fix is a standardized and independent re-search of the reference ensembles, because a key derived
from generator output should not be presented as ground truth.

\section{Conclusion}

We have demonstrated the risk of treating GEOM-DRUGS as a ground truth due to the evidently missing or incorrect conformers in the reference set.  These issues are downstream of a finite search that stopped too early.  \textbf{All five architectures we
tested are penalized for structures that are better than minima of released keys from benchmark on its own energy surface},
every interval excluding zero. With no
RMSD threshold and no energy window anywhere, on $15.0\%$ of
molecules a generator has already found a chemically intact structure below the \emph{entire}
released ensemble, which the reference's own unlimited-restart procedure was supposed to preclude.
Re-running that program recovers unlisted conformers on $11$ of $69$ molecules, and $107$ of $113$
audited pairs stay apart at DFT. These results show that GEOM-DRUGS precision reflects not only generator quality, but also the completeness of the reference ensemble. For future work, we consider the prevalence of this issue across all of
GEOM-DRUGS, and whether correcting the key would reorder any published ranking.
\bibliographystyle{abbrvnat}
\bibliography{refs_verifier}

\begin{thebibliography}{17}
\providecommand{\natexlab}[1]{#1}
\providecommand{\url}[1]{\texttt{#1}}
\expandafter\ifx\csname urlstyle\endcsname\relax
  \providecommand{\doi}[1]{doi: #1}\else
  \providecommand{\doi}{doi: \begingroup \urlstyle{rm}\Url}\fi

\bibitem[Axelrod and G{\'o}mez-Bombarelli(2022)]{axelrod2022geom}
S.~Axelrod and R.~G{\'o}mez-Bombarelli.
\newblock {GEOM}, energy-annotated molecular conformations for property prediction and molecular generation.
\newblock \emph{Scientific Data}, 9\penalty0 (1):\penalty0 185, 2022.

\bibitem[Bannwarth et~al.(2019)Bannwarth, Ehlert, and Grimme]{bannwarth2019gfn2}
C.~Bannwarth, S.~Ehlert, and S.~Grimme.
\newblock Gfn2-xtb---an accurate and broadly parametrized self-consistent tight-binding quantum chemical method with multipole electrostatics and density-dependent dispersion contributions.
\newblock \emph{Journal of Chemical Theory and Computation}, 15\penalty0 (3):\penalty0 1652--1671, 2019.
\newblock \doi{10.1021/acs.jctc.8b01176}.

\bibitem[Chai and Head-Gordon(2008)]{chai2008wb97xd}
J.-D. Chai and M.~Head-Gordon.
\newblock Long-range corrected hybrid density functionals with damped atom-atom dispersion corrections.
\newblock \emph{Physical Chemistry Chemical Physics}, 10\penalty0 (44):\penalty0 6615--6620, 2008.

\bibitem[Corso et~al.(2023)Corso, Xu, de~Bortoli, Barzilay, and Jaakkola]{corso2023particle}
G.~Corso, Y.~Xu, V.~de~Bortoli, R.~Barzilay, and T.~Jaakkola.
\newblock Particle guidance: non-i.i.d. diverse sampling with diffusion models.
\newblock \emph{arXiv preprint}, 2023.
\newblock arXiv:2310.13102.

\bibitem[Ganea et~al.(2021)Ganea, Pattanaik, Coley, Barzilay, Jensen, Green, and Jaakkola]{ganea2021geomol}
O.-E. Ganea, L.~Pattanaik, C.~W. Coley, R.~Barzilay, K.~F. Jensen, W.~H. Green, and T.~Jaakkola.
\newblock {GeoMol}: Torsional geometric generation of molecular 3d conformer ensembles.
\newblock In \emph{Advances in Neural Information Processing Systems (NeurIPS)}, 2021.
\newblock arXiv:2106.07802.

\bibitem[Hassan et~al.(2024)Hassan, Shenoy, Lee, Stark, Thaler, and Beaini]{hassan2024flow}
M.~Hassan, N.~Shenoy, J.~Lee, H.~Stark, S.~Thaler, and D.~Beaini.
\newblock {ET-Flow}: Equivariant flow-matching for molecular conformer generation.
\newblock In \emph{Advances in Neural Information Processing Systems (NeurIPS)}, 2024.
\newblock arXiv:2410.22388.

\bibitem[Jing et~al.(2022)Jing, Corso, Chang, Barzilay, and Jaakkola]{jing2022torsional}
B.~Jing, G.~Corso, J.~Chang, R.~Barzilay, and T.~Jaakkola.
\newblock Torsional diffusion for molecular conformer generation.
\newblock In \emph{Advances in Neural Information Processing Systems (NeurIPS)}, 2022.
\newblock arXiv:2206.01729.

\bibitem[Kynk{\"a}{\"a}nniemi et~al.(2019)Kynk{\"a}{\"a}nniemi, Karras, Laine, Lehtinen, and Aila]{kynkaanniemi2019improved}
T.~Kynk{\"a}{\"a}nniemi, T.~Karras, S.~Laine, J.~Lehtinen, and T.~Aila.
\newblock Improved precision and recall metric for assessing generative models.
\newblock In \emph{Advances in Neural Information Processing Systems (NeurIPS)}, volume~32, 2019.
\newblock arXiv:1904.06991.

\bibitem[Naeem et~al.(2020)Naeem, Oh, Uh, Choi, and Yoo]{naeem2020reliable}
M.~F. Naeem, S.~J. Oh, Y.~Uh, Y.~Choi, and J.~Yoo.
\newblock Reliable fidelity and diversity metrics for generative models.
\newblock In \emph{International Conference on Machine Learning (ICML)}, 2020.
\newblock arXiv:2002.09797.

\bibitem[Nikitin et~al.(2025)Nikitin, Dunn, Koes, and Isayev]{geomdrugs2025revisited}
F.~Nikitin, I.~Dunn, D.~R. Koes, and O.~Isayev.
\newblock Geom-drugs revisited: Toward more chemically accurate benchmarks for 3d molecule generation.
\newblock \emph{arXiv preprint}, 2025.
\newblock arXiv:2505.00169.

\bibitem[Northcutt et~al.(2021)Northcutt, Athalye, and Mueller]{northcutt2021labelerrors}
C.~G. Northcutt, A.~Athalye, and J.~Mueller.
\newblock Pervasive label errors in test sets destabilize machine learning benchmarks.
\newblock In \emph{Advances in Neural Information Processing Systems Datasets and Benchmarks Track}, 2021.

\bibitem[Pracht et~al.(2020)Pracht, Bohle, and Grimme]{pracht2020crest}
P.~Pracht, F.~Bohle, and S.~Grimme.
\newblock Automated exploration of the low-energy chemical space with fast quantum chemical methods.
\newblock \emph{Physical Chemistry Chemical Physics}, 22\penalty0 (14):\penalty0 7169--7192, 2020.

\bibitem[Pracht et~al.(2024)Pracht, Grimme, Bannwarth, Bohle, Ehlert, Feldmann, Gorges, M{\"u}ller, Neudecker, Plett, Spicher, Steinbach, Wesolowski, and Zeller]{pracht2024crest}
P.~Pracht, S.~Grimme, C.~Bannwarth, F.~Bohle, S.~Ehlert, G.~Feldmann, J.~Gorges, M.~M{\"u}ller, T.~Neudecker, C.~Plett, S.~Spicher, P.~Steinbach, P.~A. Wesolowski, and F.~Zeller.
\newblock Crest---a program for the exploration of low-energy molecular chemical space.
\newblock \emph{The Journal of Chemical Physics}, 160\penalty0 (11):\penalty0 114110, 2024.

\bibitem[{RDKit contributors}(2024)]{rdkit}
{RDKit contributors}.
\newblock {RDKit}: Open-source cheminformatics.
\newblock \url{https://www.rdkit.org}, 2024.
\newblock Automorphism-corrected RMSD via \texttt{AllChem.GetBestRMS}.

\bibitem[Sajjadi et~al.(2018)Sajjadi, Bachem, Lucic, Bousquet, and Gelly]{sajjadi2018assessing}
M.~S.~M. Sajjadi, O.~Bachem, M.~Lucic, O.~Bousquet, and S.~Gelly.
\newblock Assessing generative models via precision and recall.
\newblock In \emph{Advances in Neural Information Processing Systems (NeurIPS)}, volume~31, 2018.
\newblock arXiv:1806.00035.

\bibitem[Wang et~al.(2024)Wang, Elhag, Jaitly, Susskind, and Bautista]{wang2024mcf}
Y.~Wang, A.~A. Elhag, N.~Jaitly, J.~M. Susskind, and M.~A. Bautista.
\newblock Swallowing the bitter pill: Simplified scalable conformer generation.
\newblock In \emph{International Conference on Machine Learning (ICML)}, 2024.
\newblock MCF; arXiv:2311.17932.

\bibitem[Xu et~al.(2022)Xu, Yu, Song, Shi, Ermon, and Tang]{xu2022geodiff}
M.~Xu, L.~Yu, Y.~Song, C.~Shi, S.~Ermon, and J.~Tang.
\newblock {GeoDiff}: A geometric diffusion model for molecular conformation generation.
\newblock In \emph{International Conference on Learning Representations (ICLR)}, 2022.
\newblock arXiv:2203.02923.

\end{thebibliography}

\appendix

\section{Experimental detail}
\label{app:detail}

\begin{figure}[h]
\centering
\includegraphics[width=0.70\textwidth]{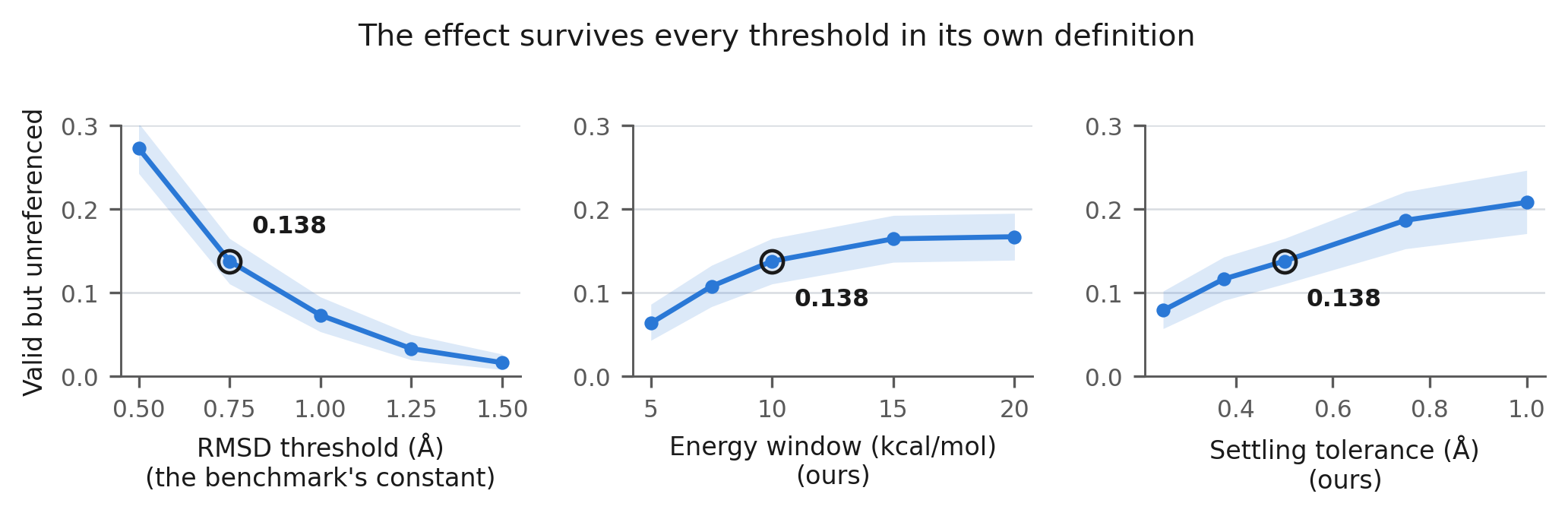}
\caption{Novel-valid fraction against each defining constant, molecule-averaged with clustered $95\%$
intervals, $n=384$, on torsional-diffusion pool and \emph{before} the reflection and
stereochemistry filters of \Cref{sec:gt}; circled point is that pool's operating point. The effect is
governed most strongly by the RMSD threshold, which is the benchmark's constant, and
tightening it raises the fraction.}
\label{fig:sens}
\end{figure}

\paragraph{Pools and protocols.}
\Cref{tab:main}, \Cref{tab:globalmin} and \Cref{tab:looarch} use public checkpoints only and
we did not retrain any model. The five checkpoints are GeoMol \citep{ganea2021geomol}, torsional
diffusion \citep{jing2022torsional}, MCF \citep{wang2024mcf}, ET-Flow \citep{hassan2024flow} and
GeoDiff \citep{xu2022geodiff}.

Three supporting measurements use a different pool, which is a particle guided
torsional-diffusion pool \citep{corso2023particle}. These three are the sensitivity sweep in \Cref{fig:sens}, the
pre-relaxation basin-crossing control and the basin characterization. We label them as such in the
text. No headline number in this paper depends on that pool.

All relaxations use xTB 6.7.1 with the GFN2-xTB method and the \texttt{--opt loose} setting.
We cap each relaxation at $500$ iterations, run one OpenMP thread per process and stop a relaxation
after $300$ seconds. The formal charge of each molecule comes from its RDKit template. All audited
molecules are closed-shell.

We report energies in kcal/mol relative to the lowest relaxed conformer of the same molecule.
Drift is the automorphism-corrected heavy-atom RMSD between the input geometry and the relaxed
geometry of the same conformer. We relax the reference conformers with the same protocol, so every
distance from a pool conformer to a reference conformer is measured between two relaxed geometries.

The DFT optimizations use GPU4PySCF and PySCF at $\omega$B97X-D3BJ/def2-SVP, with grid level
3, an SCF tolerance of $10^{-9}$, at most $100$ SCF cycles and at most $60$ geomeTRIC steps under
the GAU set. We then run a single point calculation on the optimized geometry. The driver did not
store a separate optimizer convergence flag and we computed no vibrational frequencies. For this
reason we call the results optimization endpoints and not confirmed minima.

\paragraph{The released artifact.}
We will release five things. The first is the candidate-augmented reference ensembles, which
means one key per held-out architecture together with the $n=500$ one-corrector and three-corrector
keys. We store them as pickled RDKit molecule dictionaries keyed by \texttt{corrected\_smiles}, in
the same format that \texttt{test\_mols.pkl} uses, so an existing evaluation script only needs a new
file path. The second is the per-molecule basin-count CSV files.

The third is the cached GFN2-xTB relaxation arrays for all five architectures, which hold
relaxed energies, drift values, pool-to-reference and pool-to-pool RMSD matrices, and relaxed
coordinates. These arrays are what make every number in this paper recomputable without running xTB
again. The fourth is the $69$ CREST ensembles from both settings, with their per-molecule minima,
ensemble sizes, GEOM-DRUGS keys and CREST version. The fifth is the scripts that regenerate every
table. \Cref{tab:crestkeys} gives the GEOM-DRUGS keys for the eight molecules shown in
\Cref{tab:crestfull}.

\paragraph{Basin counting.}
We group novel conformers into basins after relaxation using greedy radius clustering. The
conformer with the most neighbors within $0.75$\,\AA{} becomes a cluster center and absorbs those
neighbors. We then repeat this on the conformers that are left. This is not complete-linkage
clustering, and its output depends on the order of the conformers and on how ties are broken
whenever the $0.75$\,\AA{} graph is not transitive. We therefore treat basin counts and
candidate-key sizes as descriptive numbers. The fractions that we report at the conformer level use
no clustering across conformers, so they are not affected by this.

\paragraph{Statistics.}
Every interval in this paper is a bootstrap interval computed by resampling molecules and not
conformers, with $2{,}000$ to $20{,}000$ resamples. We resample molecules because the conformers
inside one molecule are strongly dependent, and resampling them one by one would make the intervals
much too narrow. We report the range that covers the middle $95\%$ of the resampled values, so an
interval does not have to be symmetric around the number that we report. Comparisons between arms
are paired on the molecule. The CREST rates use Jeffreys intervals.

We report many intervals and we do not apply a family-wise correction. Each interval is
therefore descriptive and it is not a multiplicity-controlled test. We do not treat two overlapping
intervals as evidence that two quantities are equal, and we do not treat a failure to reject a null
hypothesis as evidence for that null.

\section{Controls and characterization in full}
\label{app:controls}

\begin{table}[h]
\centering
\small
\caption{Every control, what it would have shown, and what it returned. The DFT check re-optimizes
the whole geometry at $\omega$B97X-D3BJ/def2-SVP rather than taking a single point; ``merge'' means
the novel and reference structures relax onto each other. The last two rows move \emph{against} the
deflation they test.}
\label{tab:controls}
\resizebox{\textwidth}{!}{%
\begin{tabular}{llr}
\toprule
Control & Would have shown & Result \\
\midrule
Mirror-image re-measure & enantiomers counted as novel & $33$--$65\%$ of raw removed \\
3D stereochemistry assignment & wrong stereoisomers counted & MCF $246/434$, GeoMol $0/101$ \\
Post-relaxation topology screen & rearranged species counted & retains $94.0$--$97.5\%$ of stereo/refl. survivors \\
Full DFT re-optimization & xTB-only geometric separation & $\mathbf{107/113}$ outside radius after chemical screen \\
Hessian evaluation & saddle points, not minima & $90.3\%$ novel vs $93.2\%$ matched \\
Pre-relaxation distance & relaxation manufactures novelty & $99.2\%$ already unreferenced \\
\texttt{uniqueconfs} match & key merely truncated for release & exact, $40$ of $40$ \\
Joint sweep of all constants & effect sits on a knife edge & $\ge 4.40\%$ at strictest cell \\
Window re-anchored at ref.\ min & our window is the wider one & it is narrower; fraction \emph{rises} \\
\bottomrule
\end{tabular}}
\end{table}

\paragraph{Anchoring of the energy window.}
Our retention window and the window used by CREST are measured from different zero points. We
measure our window from the lowest relaxed conformer seen for a molecule, taken from either the
generated pool or the reference ensemble. CREST measures its window from the global minimum that its
own search found. The two zero points differ exactly when a pool conformer lies below the reference
minimum. In that case our zero point is the lower of the two, so our window is narrower when it is
seen from the reference frame. We also tested the alternative, which is to anchor the window at the reference minimum. This
alternative sounds stricter, but it raises the fraction of conformers that we count. At $W=6$
kcal/mol the fraction goes from $6.45\%$ to $6.50\%$ for torsional diffusion and from $18.25\%$ to
$19.00\%$ for MCF.

\section{An electronic-energy weight proxy}
\label{app:boltzmann}

\paragraph{Definition of the proxy weight.}
A conformer either is or is not in the key, so a count alone cannot tell us whether a missing
conformer is low in energy or sits at the top of the retention window. We therefore compute a weight
for each conformer. The weight is $e^{-\Delta E/kT}$ with $kT=0.5925$ kcal/mol at $298$ K, where
$\Delta E$ is the energy of that conformer above the lowest energy conformer of the same molecule.
The weight falls quickly with energy. A conformer $2$ kcal/mol above the lowest one carries about
$3\%$ of the weight, and a conformer at the top of a $6$ kcal/mol window carries about
$4\times10^{-5}$.

We compute the proxy for one molecule in three steps. We take the conformers that passed the
reflection and stereochemistry screens. We cluster them at $0.75$\,\AA{} and keep the lowest energy
conformer of each cluster. We then report $\sum_{\mathrm{missing}} e^{-E/kT} / (\sum_{\mathrm{ref}}
e^{-E/kT} + \sum_{\mathrm{missing}} e^{-E/kT})$ using the cached GFN2-xTB energies. No new quantum
chemistry is involved. This quantity is not a thermodynamic population and it is not a free energy. It leaves out
zero-point, thermal, vibrational and rotational terms, and it also leaves out symmetry, degeneracy,
solvent and basin volume. It clusters the missing conformers while leaving the released conformers
unclustered. It was computed before we added the post-relaxation topology screen. These differences
can move the number in either direction, so we use it only to show where candidate missing
conformers sit on the electronic-energy scale.

\paragraph{Distribution across molecules.}
For each molecule we take the largest value among the five architectures. The mean omitted
weight over molecules is $8.97\%$ $[6.53, 11.59]$ at $W=6$ kcal/mol and $9.23\%$ $[6.76, 11.89]$ at
$W=10$ kcal/mol. The median is $0.04\%$. The mean and the median are far apart because the quantity
is bimodal. Most molecules have almost no omitted weight, and a small number of molecules have a
large amount. The next four percentages are nested, so each group contains the one after it. Of all
molecules, $27.3\%$ assign more than $1\%$ of the proxy weight to missing conformers, $16.1\%$ assign
more than $5\%$, $14.1\%$ assign more than $10\%$ and $7.3\%$ assign more than half. The mean value per architecture is $1.68\%$ for GeoMol, $3.93\%$ for GeoDiff, $4.14\%$ for
torsional diffusion, $4.17\%$ for MCF and $5.05\%$ for ET-Flow. This ordering follows the ordering in
\Cref{tab:main}.

Some molecules have a missing conformer that is lower in energy than their own reference
minimum. Across the five architectures this is $7.9\%$ to $15.4\%$ of the molecules that have any
missing conformer. In those molecules the missing conformer carries a median of $88\%$ to $100\%$ of
the proxy weight, which places it on the low energy side of a barrier. Across the same five
architectures, $5.9\%$ to $9.7\%$ of screened candidate conformers lie within $1$ kcal/mol of the
global minimum of their molecule, and $24.7\%$ to $39.2\%$ lie within $3$ kcal/mol. The mean is
therefore driven by a small group of low-lying missing conformers.

\section{The L1 rotamer at DFT}
\label{app:l1dft}

The $4.85$ kcal/mol gap in \Cref{fig:l1} is measured with GFN2-xTB, and GFN2-xTB is known to
be unreliable for amide rotational preferences in conjugated hydrazones. We therefore re-optimized a
set of conformers at $\omega$B97X-D3BJ/def2-SVP, using the same code path and the same convergence
settings as the basin-distinctness audit in \Cref{app:controls}. We took the ten lowest energy
conformers of each rotamer that are at least $0.75$\,\AA{} apart from each other, so that the ten
conformers on each side represent ten different basins and not ten copies of one basin. We added the
three lowest released reference conformers. This gives $23$ full geometry optimizations.

We ran the control first. A conformer that starts on the syn side and relaxes to the anti
side during the optimization no longer measures the syn side. None of the conformers did this. All
$23$ kept their rotamer, and the median movement away from the GFN2-xTB geometry was $0.483$\,\AA{},
with a maximum of $1.445$\,\AA{}. The energy gap survives at this higher level of theory. The rotamer of the released ensemble sits $1.908$ kcal/mol above the rotamer that CREST reaches, compared with $4.847$ kcal/mol on GFN2-xTB. The three released reference conformers sit $2.313$ to $3.099$ kcal/mol above the lowest DFT conformer. GFN2-xTB therefore gives a gap that is $2.9$ kcal/mol larger than the DFT gap, which
is a large difference in magnitude. If we assume equal partition functions for the two rotamers, then a gap of $1.908$ kcal/mol gives the released rotamer a two-state electronic-energy weight of $3.8\%$. We do not read this number as an equilibrium population. The conclusion that we rely on is that the missing rotamer is
still lower in energy after DFT optimization, and that none of the $23$ conformers changed rotamer
during that optimization.

\section{Independent validation of the oracle}
\label{app:trust}

\paragraph{Validation on two experimental sets.}
We tested our criterion on two sets of experimentally determined structures. One set should
pass the criterion and the other set should fail it. The first set is $60$ organic small-molecule
crystal structures from the Cambridge Structural Database, restricted to experimentally determined
entries. For each entry we rebuilt the heavy-atom geometry of one molecule, added hydrogen atoms,
and ran the same protocol as everywhere else in this paper, with no reference ensemble involved. A real crystal structure should already sit close to a minimum, and it does. The gradient
norm at the raw geometry is $0.116$ $[0.105, 0.131]$ $E_h/a_0$, and relaxation moves the geometry by
only $0.197$ $[0.167, 0.230]$\,\AA{}. The second set is $40$ bioactive ligands from the PDB Chemical Component Dictionary. These structures are equally real, but a protein pocket holds them in place, so they are not gas-phase minima. Our criterion marks them as strained, with a gradient norm of $0.277$ $[0.240, 0.317]$ and a drift of $0.605$ $[0.486, 0.734]$\,\AA{}. The difference between the two sets survives when we
restrict both sets to a common window of $15$ to $40$ heavy atoms and drop every phosphate-bearing cofactor, which gives $0.211$ $[0.178, 0.248]$ for the ligands against $0.117$ $[0.104, 0.131]$ for
the crystal structures. Those two intervals do not overlap. This is the known ligand-strain effect, and it marks the scope of our criterion. The criterion describes isolated-molecule physics. It should not be applied to conformers whose geometry is set by an environment. \Cref{tab:energyladder} places the generated conformers between the two experimental sets on quantities that are normalized for molecule size.

\begin{table}[h]
\centering
\small
\caption{Energy-surface quantities, \textbf{size-normalized and size-matched} ($\Delta E$ per heavy
atom, $\|\nabla V\|$ per $\sqrt{N}$, common $18$--$32$ heavy-atom window; ordering unchanged from raw
values). Adjacent intervals touch, so the ordering is the claim, not each gap. The two experimental
sets \emph{bracket} the generated ones. A crystal geometry is not a gas-phase minimum either, it is
packing-distorted with idealized hydrogens, so the CSD row is a conservative upper bound.}
\label{tab:energyladder}
\resizebox{\textwidth}{!}{%
\begin{tabular}{lrrrr}
\toprule
Population & $N$ & $\|\nabla V\|/\sqrt{N}$ & $\Delta E/N$ & drift (\AA{}) \\
\midrule
CSD small-molecule crystals (exp.) & $41$ & $0.0238$ $[0.0207, 0.0282]$ & $0.63$ $[0.49, 0.81]$ & $0.197$ $[0.167, 0.230]$ \\
Generated, settled low-energy & $256$ & $0.0290$ $[0.0274, 0.0308]$ & $1.25$ $[1.10, 1.43]$ & --- \\
Generated, strained & $44$ & $0.0368$ $[0.0305, 0.0438]$ & $2.60$ $[1.96, 3.37]$ & --- \\
PDB protein-bound ligands (exp.) & $20$ & $0.0536$ $[0.0462, 0.0615]$ & $2.76$ $[2.15, 3.53]$ & $0.605$ $[0.486, 0.734]$ \\
\bottomrule
\end{tabular}}
\end{table}

\paragraph{Noise floor of the topology perceiver.}
We perceive connectivity and stereochemistry from 3D coordinates alone and compare the
resulting canonical SMILES with the target SMILES. We did this for the raw geometry and the relaxed
geometry of $830$ conformers, drawn from $83$ molecules. The raw pass measures how often the
perceiver itself is correct, which is $91.6\%$ $[86.1, 96.3]$. This is why the topology filter of
\Cref{tab:globalmin} has to be read as a screen and not as exact chemical perception. Among the conformers that the perceiver reads correctly at the input geometry, the bond graph survives relaxation in $96.6\%$ $[92.1, 100]$ of cases. Without that condition the figure is $89.8\%$. The difference between $91.6\%$ and $89.8\%$ is $1.8$ points, which is what our relaxation pipeline costs on top of the perceiver's own error rate, and it comes entirely from $3$ molecules that carry a protonated amine center. On the two experimental sets the check is exact, because raw and relaxed perception agree row by row on all $60$ crystal structures and all $40$ PDB ligands.

\paragraph{Comparison with a higher level of theory.}
We ran single point calculations at $\omega$B97X-D3BJ/6-31G* in pyscf on $30$ relaxed
conformers. We use this functional because the original $\omega$B97X-D functional of
\citet{chai2008wb97xd} is not implemented in the dispersion module of pyscf, so we report the closest
supported member of the same family instead of substituting one silently. Over $17$ of these
conformers, the within-molecule Spearman rank correlation with the GFN2-xTB relative energies is
$0.9355$. A properly paired batch of six molecules, each contributing $3$ valid and $3$ strained
conformers, gives $0.8762$ over all $36$ conformers. This means that the cheap method orders the conformers of a molecule in essentially the same way as the expensive one. Ordering is a weaker statement than basin distinctness. \Cref{sec:gt} reports the distinctness check itself, which uses full DFT optimization instead of a single point.

\paragraph{Effect of which geometry is scored.}
We also asked whether strained conformers sit higher on the DFT surface. When we score the
xTB-relaxed geometries, the difference is small and has the wrong sign, with a median of $-0.55$
kcal/mol. This is an artifact of how the test was built, because the strained label describes the
raw geometry and relaxation removes the very difference that we are trying to measure. When we re-score the unrelaxed geometries of the same conformers, the sign reverses. Strained
conformers sit $20.35$ kcal/mol $[+0.68, +60.25]$ above valid ones, and the difference is positive in
$5$ of the $6$ molecules. The interval is wide because the unit of analysis is six molecules, and we
predicted the direction before running the test. The useful part of this result is the reversal
itself. The same comparison returns $-0.55$ or $+20.35$ depending only on which geometry is scored,
which is a warning about protocol and not a result about conformers.

\section{The global-minimum ladder}
\label{app:ladder}

\begin{table}[h]
\centering
\small
\caption{How often a generator finds a relaxed structure below the released reference minimum
($n=100$ molecules, molecule-bootstrapped $95\%$ intervals; GFN2-xTB, no RMSD threshold and no energy
window anywhere). ``Valence signature'' requires the winning relaxed structure to retain the intended
element/neighbor/H-count multiset.}
\label{tab:globalmin}
\begin{tabular}{lrl}
\toprule
Requirement & Rate & $95\%$ interval \\
\midrule
some generator's pool minimum is lower & $22.0\%$ & $[14.0, 30.0]$ \\
\quad + margin $>0.5$\,kcal/mol & $16.0\%$ & $[9.0, 23.0]$ \\
\quad + winning structure keeps the valence signature & $\mathbf{15.0\%}$ & $[8.0, 22.0]$ \\
\bottomrule
\end{tabular}
\end{table}

\begin{table}[h]
\centering
\small
\caption{Eight of the $69$ CREST molecules, chosen to span the full range of outcomes rather
than to summarize it: the six largest margins, one sub-threshold case and one null on a large
reference. Positive means CREST reached \emph{below} the released reference minimum (kcal/mol, our
protocol, GFN2-xTB). \textbf{Bold} marks a favorable margin above the prespecified
$0.5$\,kcal/mol numerical threshold; $^{\dagger}$ marks an outcome not counted as such an
improvement, either because it is smaller or because CREST is higher. ``Gain'' is what the
$4\times$ budget bought over the default setting. Campaign-level rates over all $69$ are in
\Cref{sec:crest}.}
\label{tab:crestfull}
\footnotesize
\begin{tabular}{lrrrrrrr}
\toprule
Mol & $n_{\mathrm{ref}}$ & heavy & $n_{\mathrm{rot}}$ & confs $1\times/4\times$
  & $1\times$ vs.\ ref & $4\times$ vs.\ ref & gain \\
\midrule
\multicolumn{8}{l}{\emph{CREST reaches below the released minimum at the default setting}} \\
L1 & $135$ & $22$ & $7$ & $357\,/\,661$ & $\mathbf{+4.847}$ & $\mathbf{+4.847}$ & $0.000$ \\
T13 & $8$ & $22$ & $3$ & $10\,/\,13$ & $\mathbf{+2.017}$ & $\mathbf{+2.017}$ & $0.000$ \\
\midrule
\multicolumn{8}{l}{\emph{the $4\times$ budget is what buys it}} \\
L3 & $30$ & $20$ & $4$ & $107\,/\,319$ & $0.000$ & $\mathbf{+2.742}$ & $+2.742$ \\
Y1 & $5$ & $18$ & $3$ & $10\,/\,5$ & $0.000$ & $\mathbf{+1.882}$ & $+1.882$ \\
U5 & $64$ & $26$ & $5$ & $161\,/\,390$ & $0.000$ & $\mathbf{+1.301}$ & $+1.301$ \\
\midrule
\multicolumn{8}{l}{\emph{the released search was better, until the budget rose}} \\
molA & $18$ & $21$ & $3$ & --- & $-0.524^{\dagger}$ & $\mathbf{+1.483}$ & $+2.007$ \\
\midrule
\multicolumn{8}{l}{\emph{below the numerical threshold, and a null}} \\
D5 & $24$ & $26$ & $5$ & $521\,/\,827$ & $+0.273^{\dagger}$ & $+0.242^{\dagger}$ & $-0.031$ \\
M2 & $196$ & $24$ & $7$ & $277\,/\,431$ & $0.000$ & $0.000$ & $0.000$ \\
\bottomrule
\end{tabular}
\end{table}

\begin{table}[h]
\centering
\footnotesize
\caption{GEOM-DRUGS keys (\texttt{corrected\_smiles}) and formal charges for the molecules of
\Cref{tab:crestfull}, so each row is resolvable to a GEOM entry. Keys for all $69$ ship with the
released artifact.}
\label{tab:crestkeys}
\scriptsize
\begin{tabular}{llr}
\toprule
Tag & GEOM-DRUGS key & charge \\
\midrule
\texttt{D5} & \texttt{CCc1nc([NH2+]c2cc(OC)ccc2OC)c2oc3ccccc3c2n1} & $1$ \\
\texttt{L1} & \texttt{C=CCOc1ccccc1/C=N/NC(=O)Cc1ccccc1} & $0$ \\
\texttt{L3} & \texttt{C[C@@H](C)NC(=O)C(=O)NNC(=O)CCc1ccccc1} & $0$ \\
\texttt{M2} & \texttt{C=CCN(c1ccccc1C(=O)O)S(=O)(=O)/C=C/c1ccccc1} & $0$ \\
\texttt{T13} & \texttt{O=C(N/N=C/c1c(O)ccc2ccccc12)c1[nH]ncc1Br} & $0$ \\
\texttt{U5} & \texttt{COc1cc2c(C(=O)NCc3ccccn3)cn(C)c(=O)c2cc1OC} & $0$ \\
\texttt{Y1} & \texttt{Cc1cccc(/C=N/NC(=O)c2ccncc2)n1} & $0$ \\
\texttt{molA} & \texttt{CC(=O)N1CC[NH+](CC(=O)Nc2c(Cl)cccc2Cl)CC1} & $1$ \\
\bottomrule
\end{tabular}
\end{table}

\section{The exploratory PDB screen}
\label{app:pdb}

The initial screen contains $14$ rows that match an entry in the PDB Chemical Component
Dictionary, and those $14$ rows correspond to $13$ unique GEOM-DRUGS keys. For each row the screen
measures two distances to the same experimental geometry. The first is $d(R,P)$, from the released
reference ensemble to the PDB geometry. The second is $d(g,P)$, from the closest conformer of each
generator to the PDB geometry. Two things limit what this screen can show. Eight of the $14$ matches required a fallback that strips stereochemistry from the graph before matching. The generated pools also hold many more conformers than the reference ensembles. The screen is therefore useful for proposing candidate examples, and it cannot declare a precision failure on its own. A precision claim also needs exact
chemical identity and a direct measurement of $d(g,R)$, which is the distance from the selected generated conformer to every released reference conformer. The body reports the two examples that survive those checks.

\begin{table}[h]
\centering
\small
\caption{Exploratory two-leg screen of matched PDB bound ligands, per architecture.
$d_{\mathrm{ref}}$ is the distance from the \emph{entire} released reference ensemble to the
experimental geometry; the five middle columns are each architecture's own closest conformer, bold
when inside the $0.75$\,\AA{} radius, and the last column counts how many architectures are.
``\,---\,'' means that pool does not cover the molecule. Pool depths are ${\sim}8n_{\mathrm{ref}}$
(GeoMol capped at $400$, GeoDiff at $30$). These two-leg distances do not by themselves determine
whether a generated conformer fails precision against the reference set.}
\label{tab:pdbfull}
\resizebox{\textwidth}{!}{%
\begin{tabular}{llrrrrrrrr}
\toprule
Ligand & PDB & $n_{\mathrm{ref}}$ & $d_{\mathrm{ref}}$ & GeoMol & Tors.\ diff. & MCF & ET-Flow & GeoDiff & $\#{<}0.75$ \\
\midrule
\multicolumn{10}{l}{\emph{candidates from the two-leg screen; direct checks reported in the body}} \\
\midrule
\textbf{WNA} & 5RWU & $30$ & $\mathbf{0.800}$ & $\mathbf{0.276}$ & $\mathbf{0.656}$ & $\mathbf{0.254}$ & $\mathbf{0.286}$ & $\mathbf{0.642}$ & $\mathbf{5}$ \\
\textbf{3VC} & 4WWN & $22$ & $\mathbf{0.872}$ & $\mathbf{0.702}$ & $\mathbf{0.698}$ & $\mathbf{0.605}$ & $\mathbf{0.223}$ & $1.075$ & $\mathbf{4}$ \\
\textbf{8DE} & 7W7F & $55$ & $\mathbf{0.849}$ & $\mathbf{0.562}$ & $0.795$ & $\mathbf{0.615}$ & $\mathbf{0.463}$ & $1.489$ & $\mathbf{3}$ \\
\textbf{T7J} & 5RFM & $6$ & $\mathbf{0.932}$ & $1.047$ & $1.045$ & $\mathbf{0.447}$ & $1.216$ & $1.000$ & $\mathbf{1}$ \\
\midrule
\multicolumn{10}{l}{\emph{all remaining molecules}} \\
\midrule
SDL & 5GMG & $150$ & $0.281$ & $0.815$ & --- & $\mathbf{0.275}$ & $\mathbf{0.253}$ & $\mathbf{0.701}$ & $3$ \\
SMN & 1MDL & $4$ & $0.306$ & $\mathbf{0.142}$ & $\mathbf{0.303}$ & $\mathbf{0.176}$ & $\mathbf{0.298}$ & $\mathbf{0.430}$ & $5$ \\
0RF & 4EKL & $530$ & $0.532$ & $1.346$ & $0.781$ & $\mathbf{0.551}$ & $\mathbf{0.556}$ & $1.663$ & $2$ \\
1U3 & 4KZ5 & $37$ & $0.657$ & $\mathbf{0.386}$ & $\mathbf{0.332}$ & $\mathbf{0.347}$ & $\mathbf{0.302}$ & $0.876$ & $4$ \\
RMN & 1MDL & $4$ & $0.715$ & $\mathbf{0.393}$ & $\mathbf{0.734}$ & $\mathbf{0.198}$ & $\mathbf{0.625}$ & $\mathbf{0.379}$ & $5$ \\
JDC & 4DJH & $53$ & $1.173$ & $1.293$ & $1.112$ & $1.003$ & $2.309$ & $1.905$ & $0$ \\
RXM & 6BEE & $94$ & $1.266$ & $2.009$ & --- & $1.468$ & $1.524$ & $1.704$ & $0$ \\
LBL & 6VTO & $161$ & $1.668$ & $1.704$ & $0.971$ & $1.000$ & $1.009$ & $1.490$ & $0$ \\
EN9 & 6LQG & $131$ & $2.113$ & $1.375$ & $0.855$ & $1.000$ & $1.168$ & $1.724$ & $0$ \\
TQI & 2Y6C & $18$ & $2.192$ & $0.846$ & $1.195$ & $1.230$ & $1.029$ & $1.513$ & $0$ \\
\bottomrule
\end{tabular}}
\end{table}

\paragraph{Direct audit of the four screened candidates.}
The screen proposed four candidates and we checked each one directly. For 3VC the selected
ET-Flow conformer has $d(g,P)=0.223$\,\AA{} and $d(g,R)=0.847$\,\AA{}. For 8DE the same two values
are $0.463$\,\AA{} and $0.772$\,\AA{}. Both of these are precision failures with exact
stereochemistry. The WNA conformer from MCF has $d(g,R)=0.713$\,\AA{}, which is inside the
$0.75$\,\AA{} radius, so it passes the benchmark. We removed T7J because its fallback match has the
opposite configured stereocenter from the GEOM-DRUGS key. This third distance is why the body
reports two examples and not the four candidates that the screen proposed. A future quantitative study should match protomer, tautomer and stereochemistry exactly, and should use repeated subsampling at the benchmark's own budget of $2n_{\mathrm{ref}}$ conformers, or recovery curves.

\paragraph{Scope.}
Conformers bound inside a protein pocket can be strained, so this section is not evidence
about gas-phase completeness. It shows one possible downstream consequence for scoring, and it
motivates a properly powered benchmark based on experimental overlap. Six further GEOM-DRUGS molecules have crystal structures in the Cambridge Structural Database, covering $15$ determinations, with released ensembles of $1$ to $16$ conformers. The released ensemble misses the crystal conformer on $2$ of these $6$ molecules, and a generator is closer to the crystal conformer than the whole released ensemble on $10$ of the $15$ determinations. No case in this set has a generator inside the $0.75$\,\AA{} radius where the released ensemble is outside it, which is why we lead with the PDB ligands.

\end{document}